\documentclass[final,5p,times,authoryear]{elsarticle}
\usepackage{amssymb}
\journal{Computerized Medical Imaging and Graphics}

\usepackage[
pdfauthor={jdq},
bookmarks=true,
colorlinks=true,
linkcolor=blue,
urlcolor=blue,
citecolor=blue,
linktocpage=true,   % makes the page number as hyperlink in table of content
hyperindex=true
]{hyperref}
\usepackage{amsmath}
\usepackage{amssymb}
\usepackage{verbatim}
\usepackage{multirow}
\usepackage{color,graphicx}
\usepackage{arydshln}
\usepackage{esvect}
\usepackage{subfigure}
\usepackage[]{caption} 
\usepackage[switch]{lineno}
\usepackage[compress]{cite}
\usepackage{booktabs}
\usepackage{adjustbox}
\usepackage[figuresright]{rotating}
\usepackage{cleveref}
\usepackage{threeparttable}
\providecommand{\Zxhreftb}[1]{Table~\ref{#1}}
\providecommand{\Zxhreftbs}[1]{Tables~\ref{#1}}

\providecommand{\Zxhreffig}[1]{Fig.~\ref{#1}}

\newcommand{\tabincell}[2]{\begin{tabular}{@{}#1@{}}#2\end{tabular}}

\begin{document}
\captionsetup[figure]{labelfont={bf},labelformat={default},labelsep=period,name={Fig.}}
\captionsetup[table]{labelfont={bf},labelformat={default},labelsep=newline,name={Table}}
\biboptions{numbers,sort&compress}
\begin{frontmatter}

\title{Learning-Based Algorithms for Vessel Tracking: A Review}

\author[1]{Dengqiang~Jia}
\author[2]{Xiahai~Zhuang \corref{mycorrespondingauthor}}\ead{zxh@fudan.edu.cn}
\cortext[mycorrespondingauthor]{Corresponding author}

\address[1]{School of Naval Architecture, Ocean and Civil Engineering, Shanghai Jiao Tong University, Shanghai, China}
\address[2]{School of Data Science, Fudan University, Shanghai, China}

\begin{abstract}
Developing efficient vessel-tracking algorithms is crucial for imaging-based diagnosis and treatment of vascular diseases.
Vessel tracking aims to solve recognition problems such as key (seed) point detection, centerline extraction, and vascular segmentation. Extensive image-processing techniques have been developed to overcome the problems of vessel tracking that are mainly attributed to the complex morphologies of vessels and image characteristics of angiography.
This paper presents a literature review on vessel-tracking methods, focusing on machine-learning-based methods.
First, the conventional machine-learning-based algorithms are reviewed, and then, a general survey of deep-learning-based frameworks is provided.
On the basis of the reviewed methods, the evaluation issues are introduced.
The paper is concluded with discussions about the remaining exigencies and future research.

\end{abstract}

\begin{keyword}
Vessel tracking, Learning-based algorithms, Review

\end{keyword}

\end{frontmatter}
\section{Introduction}\label{Introduction}
Blood vessels, spread throughout the human body, constitute a significant part of the circulatory system.
All body tissues rely on the normal functioning of different vessels such as cerebral arteries, retinal vessels, carotid arteries, pulmonary arteries, and coronaries.
Any abnormal change in or damage to the vessels will be manifested as diseases at different levels (e.g., stroke, arteriosclerosis, cardiovascular diseases, and hypertension).
Medical imaging and image analysis enable novel technologies and applications for better diagnosis and treatment of blood-vessel diseases.
Tracking the target vessels from the wide field of view of medical images is a prerequisite for the localization and identification of abnormal vessels or regions of interest.
However, manual annotation, which usually demands expertise, is particularly time-consuming and tedious.

Vessel tracking aims to solve the problems encountered in vessel image analysis, including key-point (seed point) detection, centerline extraction, and vascular segmentation. These problems considerably differ because of the wide array of vessel anatomies and image characteristics. Accordingly, the problematic factors are categorized into two groups: those related to vessel morphologies (e.g., small size, branching pattern, tortuosity, and severe stenosis of vessels) and image characteristics (e.g., low contrast, noise, artifacts, dislocation, and adjacent hyper-intense structures).

Localizing the key points and recognizing the key patterns of vascular structures are fundamental to perform vessel tracking; building models based on various assumptions on vascular appearances (i.e., the prior knowledge and intrinsic features) is also important. 
Exploring such problems facilitates the development of new algorithms, particularly in learning-based tracking methods.

In the literature, several articles focus on the survey of vessel-tracking methods. 
To the best of our knowledge, \citet{Suri2002} published the first survey on this topic. 
They concentrated on skeleton and indirect techniques for vascular segmentation. 
In addition to vessel-tracking methods, \citet{Kirbas2004a} particularly reviewed the methods of detecting similar vessel characteristics, such as neurovascular and tubular structures. 
The review of \citet{Lesage2009} further focused on lumen segmentations.
They categorized the methodologies according to three aspects (i.e., models, features, and extraction schemes) and provided general considerations for each aspects.

Recently, more survey papers for tracking the vessels of certain organs in specific imaging modality, e.g., cerebral vessel segmentation from magnetic resonance (MR) images \citep{Klepaczko2016}, coronary reconstruction from X-ray angiography \citep{Cimen2016d}, and lung vessel detection from computed tomography angiography (CT) \citep{Rudyanto2014}, have been published. 
Reviews on retinal vessel segmentation were presented in the work \citep{Abramoff2010, Fraz2012a, Singh2015, Pohankar2016, Mansour2017, LSrinidhi2017, Vostatek2017, Soomro2019b, Khan2019}. 
Particularly, \citet{Moreno2015a} were interested in formulating general methods for enhancement technologies, and \citet{Kerrien2017} focused on modeling approaches. 
In view of the potential of learning-based methods for tracking retinal vessels, \citet{Moccia2018} and  \citet{Zhao2019b} reviewed the principles and applications.
\citet{Soomro2019b} particularly focused on deep-learning-based works of retinal blood vessel segmentation.
\begin{figure*}  \center\setlength{\abovecaptionskip}{-0.cm}
\setlength{\belowcaptionskip}{-0.cm}
\includegraphics[scale=.8]{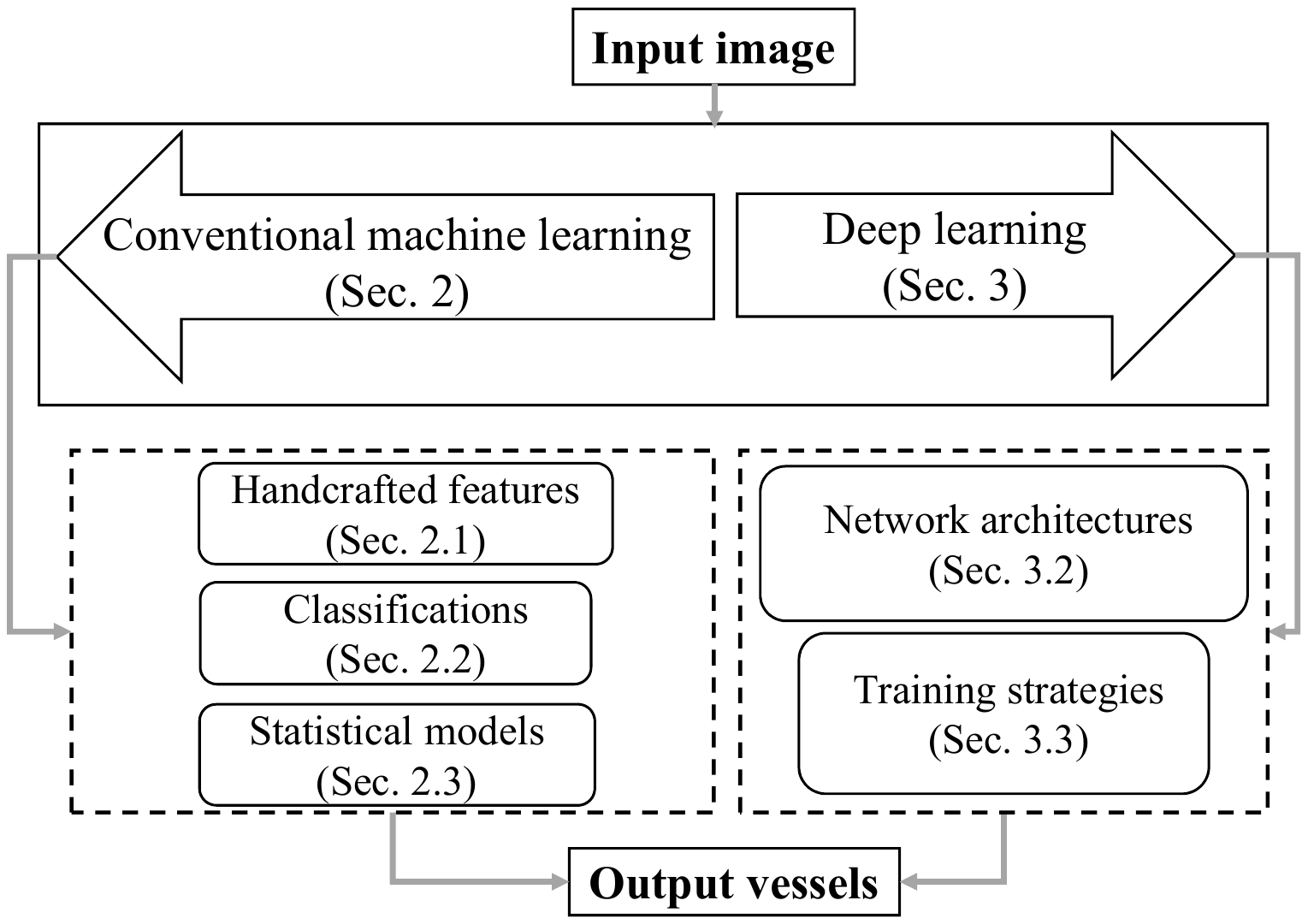} 
   \caption{Recapitulative diagram of vessel tracking using learning-based methods.}
\label{fig: processingdiagram}\end{figure*}
\vspace{-1em}

\subsection{Aims of this paper}
This work aims to provide an up-to-date review of vessel tracking based on machine-learning-methods, also referred to as learning-based methods.
We focus on the learning-based methods for tracking vessels of various organs using different imaging modalities.
The recapitulative diagram of the learning-based methods for vessel tracking is shown in \Zxhreffig{fig: processingdiagram}.
To cover the articles to a feasible extent, a search for the term vessel/vascular segmentation/extraction has been performed using engines such as PubMed \footnote{http://www.ncbi.nlm.nih.gov/pubmed}, IEEE Xplore \footnote{http://ieeexplore.ieee.org}, and Google Scholar \footnote{http://scholar.google.com}.
Among over 300 collated articles, the focus of attention was on papers published during the last 10 years.
Note that this article does not cover all the details of databases and evaluation standards that can be found in the literature \citep{Schaap2009, Hameeteman2011, Kirisli2013, Rudyanto2014, Vostatek2017, Moccia2018, Yan2018e}.

The rest of the paper is organized as follows.
Section \ref{Vessel tracking using conventional machine learning} reviews the vessel tracking methods using conventional machine learning.
Section \ref{Vessel tracking based on deep learning} reviews the vessel tracking approaches using deep learning.
Based on the reviewed methods, Section \ref{Evaluation issues} introduces the evaluation issues.
Finally, Section \ref{Conclusion and discussion} concludes the review and explores potential directions for future work on learning-based methods for vessel tracking.

\section{Vessel tracking using conventional machine learning}\label{Vessel tracking using conventional machine learning}

This section reviews the vessel-tracking works that employ conventional learning-based algorithms including the methodologies of hand-crafted features, classifications and statistical models.
\Zxhreftbs{tab:traditional learning:retinal vessel}- \ref{tab:traditional learning:other vessels} summarize the decomposition of a selection of  representative works in this field according to the applications. 
\Zxhreftbs{tab:traditional learning techniques} summarizes the existing conventional machine-learning-based works by grouping them into different subcategories.

\begin{figure*}[!t]
\centering
\setlength{\abovecaptionskip}{-0.cm}
\setlength{\belowcaptionskip}{-0.cm}
\includegraphics[scale=.5]{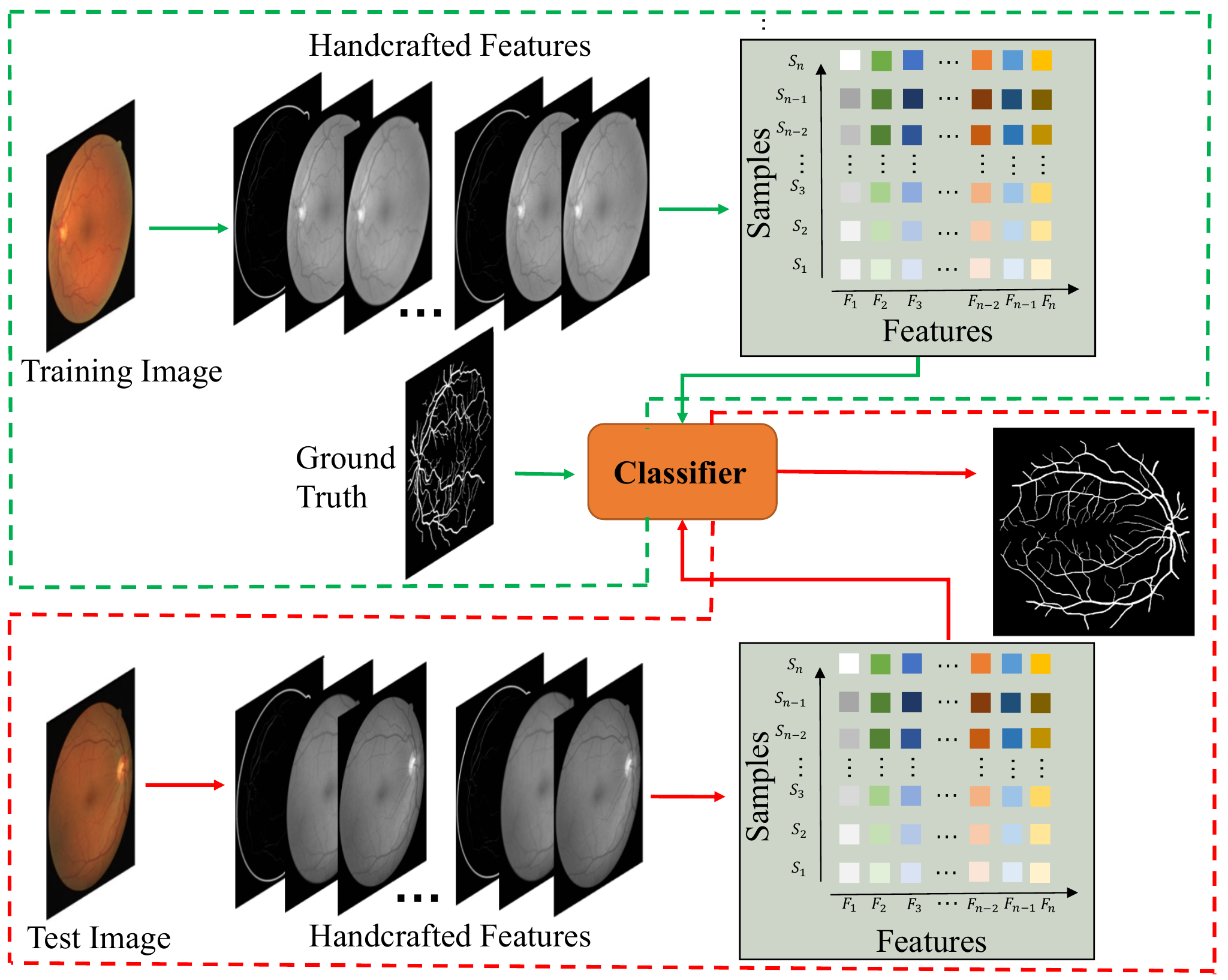}
\caption{Diagram of the retinal vessel segmentation using the conventional machine-learning method with supervised training.}\label{fig: Traditional supervised learning}
\end{figure*}
\begin{sidewaystable*}
\begin{minipage}[]{1.2\textwidth}
\caption{Overview of conventional machine-learning-based methods for tracking \textbf{retinal vessels}: the evaluation metrics and datasets are presented in Section \ref{Evaluation issues}; see list of abbreviations at the bottom.}\vspace{-0.2cm}
\label{tab:traditional learning:retinal vessel}
\centering
\scriptsize
\begin{threeparttable}
\begin{tabular}{p{2.8cm}p{3.8cm}p{4cm}p{5.5cm}p{5.2cm}}
\hline\hline
\multirow{1}{*}{Authors} &\multirow{1}{*}{Methods}&\multirow{1}{*}{Data}&\multirow{1}{*}{Experiment}&\multirow{1}{*}{Results}\\
\cmidrule(r){1-5}
\citet{Becker2013}&Boosting-based, learning kernels&Retinal colored image, DRIVE &20 for training, 20 for testing, one-off train + test&Precision-recall curves\\
\cmidrule(r){1-5}
\citet{Sironi2015}&Learning separable filters &Retinal colored image, DRIVE&20 for training, 20 for testing, one-off train + test&AUC=0.962\\
\cmidrule(r){1-5}
\multirow{2}{*}{\citet{Annunziata2016}}	&{Convolutional sparse coding-} &Retinal colored image, DRIVE&20 for training, 20 for testing , one-off train + test&AUPRC=0.87\\
{}&{filter learning}&Retinal colored image, STARE&19 for training, 1 for testing, leave-one-out&AUPRC=0.86 \\
\cmidrule(r){1-5}
\multirow{4}{*}{\citet{Gu2017}}	&\multirow{4}{*}{Boosting-based	, structured features}	&Retinal colored image, DRIVE&20 for training, 20 for testing, one-off train + test&Pr=0.7931, Re=0.7595, Sp=0.9711\\
{}&{}&Retinal colored image, STARE&20 images, five-fold cross-validation&Pr=0.7761, Re=0.7791, Sp=0.9741 \\
{}&{}&Retinal colored image, CHASE-DB1 &14 for training, 14 for testing, one-off train + test&Pr=0.6660, Re=0.6850, Sp=0.9664\\
{}&{}&Retinal colored image, HRF & half for training, half for testing, one-off train + test&Pr=0.7775, Re=0.7602, Sp=0.9795\\
\cmidrule(r){1-5}
\multirow{2}{*}{\citet{Javidi2017}}	&{Dictionary learning, } &\multirow{2}{*}{Retinal colored image, DRIVE}&\multirow{2}{*}{20 for training, 20 for testing , one-off train + test}&\multirow{2}{*}{Acc=0.9446}\\
{}&{vessel and non-vessel features}&{}&{}&{}\\
\cmidrule(r){1-5}
\multirow{2}{*}{\citet{Kalaie2017b}}	&{Hierarchical probabilistic framework, } &Retinal colored image, REVIEW&16 images, leave-one-out&Acc=0.9446\\
{}&{intensity features of the cross sections}&{Retinal colored image, DRIVE}&{40 images, leave-one-out}&{Acc=0.970}\\
\cmidrule(r){1-5}
\multirow{4}{*}{\citet{Orlando2017}}	&\multirow{4}{*}{SVM, features in CRF}	&Retinal colored image, DRIVE&20 for training, 20 for testing, one-off train + test&Pr=0.7854, Se=0.7897, Sp=0.9684\\
{}&{}&Retinal colored image, STARE&19 for training, 1 for testing, leave-one-out&Pr=0.7740, Se=0.7680, Sp=0.9738\\
{}&{}&Retinal colored image, CHASE-DB1 &8 for training, 20 for testing, one-off train + test&Pr=0.7438, Se=0.7277, Sp=0.9712\\
{}&{}&Retinal colored image, HRF & 5 for training, 40 for testing, one-off train + test&Pr=0.6950, Se=0.7794, Sp=0.9650\\	
\cmidrule(r){1-5}
\multirow{3}{*}{\citet{Zhang2017d}}	&\multirow{3}{*}{\tabincell{l}{Random forest, gaussian-based \\ filters wavelet transform}}	&Retinal colored image, DRIVE&20 for training, 20 for testing, one-off train + test&Acc=0.9466, AUC=0.9703, Se=0.7861, Sp=0.9712\\
{}&{}&Retinal colored image, STARE&19 for training, 1 for testing, leave-one-out&Acc=0.9547, AUC=0.9740, Se=0.7882, Sp=0.9729\\
{}&{}&Retinal colored image, CHASE-DB1 &27 for training, 1 for testing, leave-one-out&Acc=0.9502, AUC=0.9706, Se=0.7644, Sp=0.9716\\
\hline
\hline
\end{tabular}
\begin{tablenotes}
        \footnotesize
        \item[*] List of abbreviations: Acc=accuracy; AUC=area under the ROC curve; AUPRC=area under the precision-recall curve; CRF=conditional random field; Pr=precision;  Re=recall; REVIEW \citep{2008REVIEW}; Se=sensitivity; Sp=specificity; SVM= support vector machine.
      \end{tablenotes}
\end{threeparttable}

\end{minipage}\vspace{0.1cm}
\\
\begin{minipage}[]{1.2\textwidth}
\caption{Overview of conventional machine-learning-based methods for tracking \textbf{coronary vessels}: the evaluation metrics and datasets are presented in Section \ref{Evaluation issues}; see list of abbreviations at the bottom.}
\label{tab:traditional learning:coronary}
\vspace{-0.2cm}
\centering
\scriptsize
\begin{threeparttable}
\begin{tabular}{p{2.8cm}p{3.8cm}p{3.8cm}p{5.5cm}p{5.2cm}}
\hline\hline
\multirow{1}{*}{Authors} &\multirow{1}{*}{Methods}&\multirow{1}{*}{Data}&\multirow{1}{*}{Experiment}&\multirow{1}{*}{Results}\\
\cmidrule(r){1-5}

\multirow{2}{*}{\citet{Schaap2011}}	&Nonlinear regression, point-	&Coronary CTA, local data&82 for training, 1 for testing, leave-one-out&Distance=0.15mm\\
{}&{distribution and intensity model}&Coronary CTA, CAT08&8 for training, 24 for testing, one-off train + test&AI=0.23mm, OF=0.725, OT=0.971, OV=0.969\\
\cmidrule(r){1-5}
\multirow{2}{*}{\citet{Lesage2016}}&Bayesian vessel model and particle-  &\multirow{2}{*}{Coronary CTA, local data}&\multirow{2}{*}{10 for training, 51 for testing}&\multirow{2}{*}{AI=0.25mm, OT=0.925, OV=0.862}\\
{}&{filtering, flux-based image feature}&{}&{}\\
\cmidrule(r){1-5}
\multirow{2}{*}{\citet{Mehmet2016}}&Boosting-based, image features, &\multirow{2}{*}{Coronary CTA, local data}&\multirow{2}{*}{90 for training, 20 for testing}&\multirow{2}{*}{Se \textgreater 0.9, Sp \textgreater 0.9}\\
{}&{orientation and scale}&{}&{}\\
\hline
\hline
\end{tabular}
\begin{tablenotes}
        \footnotesize 
        \item[*] List of abbreviations: AI=average inside; CTA = computed tomography angiography; OF=overlap until first error; OT=overlap with the clinically relevant part of the vessel; OV=overlap; Se=sensitivity; Sp=specificity.
      \end{tablenotes}
\end{threeparttable}

\end{minipage}\vspace{0.1cm}
\\
\begin{minipage}[]{1.2\textwidth}
\caption{Overview of conventional machine-learning-based methods for tracking \textbf{other vessels}: the evaluation metrics and datasets are presented in Section \ref{Evaluation issues}; see list of abbreviations at the bottom.}
\label{tab:traditional learning:other vessels}\vspace{-0.2cm}
\centering
\scriptsize
\begin{threeparttable}
\begin{tabular}{p{2.8cm}p{3.8cm}p{3.8cm}p{5.5cm}p{5.4cm}}
\hline\hline
\multirow{1}{*}{Authors} &\multirow{1}{*}{Methods}&\multirow{1}{*}{Data}&\multirow{1}{*}{Experiment}&\multirow{1}{*}{Results}\\
\cmidrule(r){1-5}
\multirow{2}{*}{\citet{Bogunovic2012a}}&\multirow{2}{*}{SVM, bifurcation features} &Internal carotid artery 3DRA, &\multirow{2}{*}{96 images, cross-validation}&\multirow{2}{*}{Cross-validation success rate =0.99}\\
{}&{}&{local data}&{}\\
\cmidrule(r){1-5}
\multirow{2}{*}{\citet{Cheng2012a}}&\multirow{2}{*}{Seed searching}&\multirow{2}{*}{Mammography, local data}&1800 samples, 1200 for training, 400 for testing, &\multirow{2}{*}{Se=0.93, Sp=0.851}\\
{}&{}&{}&{four-fold cross-validation}&{}\\
\cmidrule(r){1-5}
\multirow{2}{*}{\citet{Zheng2012a}}&Non-rigid deformation, position,&\multirow{2}{*}{Aorta C-arm CT, local data}&\multirow{2}{*}{319 volumes, four-fold cross-validation}&\multirow{2}{*}{Mean error=1.08 mm}\\
{}&{orientation and scale}&{}&{}\\
\cmidrule(r){1-5}
\multirow{2}{*}{\citet{Cherry2015}}&Random forest, intensity, vesselness, &\multirow{2}{*}{Pelvis CT angiograms, local data}&\multirow{2}{*}{10 for training, 30 for testing, one-off train + test}&\multirow{2}{*}{Pr=0.752, Re=0.677}\\
{}&{ray-casting, MIP, and spanning tree}&{}&{}\\
\cmidrule(r){1-5}
\multirow{2}{*}{\citet{Rempfler2015b}}&Probabilistic model, &\multirow{2}{*}{Mouse brain MR images, local data}&\multirow{2}{*}{4 for training, 1 for testing, leave-one-out}&\multirow{2}{*}{DSC=0.516}\\
{}&{physiological-geometric properties}&{}&{}\\
\cmidrule(r){1-5}
\citet{Schneider2015a}&Random forest, oriented features &Synthetic vascular data, local data&4 for testing, leave-one-out&DSC=0.95\\
\cmidrule(r){1-5}
\multirow{2}{*}{\citet{Zhang2017g}}	&Random forest, steerable- &\multirow{2}{*}{Perivascular 7T MR, local data}&\multirow{2}{*}{19 image sets, two-fold cross-validation}&\multirow{2}{*}{DSC=0.661, Se=0.651}\\
{}&{frangi-filters and OOF}&{}&{}\\
\cmidrule(r){1-5}
\citet{Lorza2018}&SVM, a radial basis function kernel &Carotid bifurcation MRI, local data&49 arteries for testing&DSC wall overlap= 0.741\\
\hline
\hline
\end{tabular}
\begin{tablenotes}
        \footnotesize
        \item[*] List of abbreviations: 3DRA=3D rotational angiography; CT=computed tomography; DSC=dice similarity coefficient; MIP=maximum intensity projection; MR=magnetic resonance; MRI=magnetic resonance imaging; OOF=optimally oriented flux; Pr=precision; Re=recall; Se=sensitivity; Sp=specificity.
      \end{tablenotes}
\end{threeparttable}

\end{minipage}

\end{sidewaystable*}

\begin{sidewaystable*}
\vspace{0.8cm}
 \begin{minipage}[t]{1.1\textwidth}\vspace{-0.8cm}
  \centering
\caption{Overview of the techniques in traditional machine-learning-based methods: see list of abbreviations at the bottom. }
\label{tab:traditional learning techniques}\vspace{-0.2cm}
\begin{threeparttable}
           \begin{tabular}{p{5cm}p{4cm}p{13cm}}
\hline\hline
\multirow{1}{*}{Conventinal learning-based methods} &\multirow{1}{*}{Techniques}&\multirow{1}{*}{}\\
\cmidrule(r){1-3}
\multirow{9}{*}{Hand-crafted featues}	&Intensity features:& \citet{Vukadinovic2010, Cherry2015, Mehmet2016}\\
\cdashline{2-3}[0.8pt/2pt]
{}&Intensity gradient features: &\citet{Mehmet2016}\\
\cdashline{2-3}[0.8pt/2pt]
{}&Bifurcation feature vectors: &\citet{Bogunovic2012a}\\
\cdashline{2-3}[0.8pt/2pt]
{}&Spatial location: &\citet{Vukadinovic2010, Mehmet2016}\\
\cdashline{2-3}[0.8pt/2pt]
{}&Angles: &\citet{Bogunovic2012a, Mehmet2016}\\
\cmidrule(r){2-3}
{}&Learning-based kernels: &\citet{Poletti2014c, Liu2014c, Lesage2016, Asl2017}\\
\cdashline{2-3}[0.8pt/2pt]
\multirow{2}{*}{}&\multirow{2}{*}{Learning-based filters:} &\citet{Lin2012b, Azzopardi2013c, Annunziata2015a, Sironi2015}\\ 
{}&{}&\citet{Annunziata2016, Zhang2017g, Deng2018, Javidi2017}\\
\cmidrule(r){1-3}
\multirow{12}{*}{Classifications}&\multirow{2}{*}{K-means: }&\citet{Coates2012, Saffarzadeh2014, Zhang2014d, Jodas2017}\\ {}{}{}&{}&\citet{Goceri2017a, Lu2017a, Xia2018}\\
\cmidrule(r){2-3}
{}&\multirow{2}{*}{Fuzzy C-means clustering:} &\citet{Mapayi2015, Mapayi2016, Khan2016}\\
{}&{}&\citet{Haddad2018, Zeng2018b}\\
\cmidrule(r){2-3}
{}&\multirow{2}{*}{Support Vector Machine:} &\citet{A.Osareh2009, You2011, Hanaoka2015, Chen2015c}\\
{}&{}&\citet{Kang2015a, Jawaid2017b, Orlando2017, Lorza2018}\\
\cmidrule(r){2-3}
{}&\multirow{3}{*}{Boosting-based methods:} &\citet{Turetken2016, Lupascu2010, Gu2017}\\
{}&{}&\citet{Memari2017, Lupascu2013e}\\
{}&{}&\citet{Fraz2012b, Lupascu2013e, Hashemzadeh2019}\\
\cmidrule(r){2-3}
{}&\multirow{2}{*}{Random forest:} &\citet{Annunziata2015, Melki2014, Zhang2017d}\\
{}&{}&\citet{Schneider2015a, Sankaran2016, Cherry2015}\\
\cmidrule(r){2-3}
{}&\multirow{1}{*}{Hybrid classifiers:} &\citet{Rani2016, Lugauer2014a, Chapman2015, Hu2018e}\\
\cmidrule(r){1-3}
\multirow{5}{*}{Statistical models}&\multirow{1}{*}{Threshold: }&\citet{Vukadinovic2010, Cheng2012a}\\ 
\cmidrule(r){2-3}
{}&\multirow{1}{*}{Intensity/appearance model:} &\citet{Schaap2011, Zheng2012a, Rempfler2014a}\\
\cmidrule(r){2-3} 
{}&\multirow{2}{*}{Topological model:} &\citet{Rempfler2015b, Asl2017, Zhao2017d}\\ 
{}&{}&\citet{Chai2013, Kalaie2017b}\\

\hline
\hline
\end{tabular}
\end{threeparttable}
 \end{minipage}
    \end{sidewaystable*}

\subsection{Hand-crafted features}\label{hand-crafted features}
A broad definition of hand-crafted features is provided in \citep{Lesage2009}.
Conventional machine-learning-based methods train models with numerous hand-crafted features, which should be well-designed according to the applications.
These features (i.e., global and local features) can be obtained by a series of filters such as those given in \citep{Frangi1998, Agam2005a, Manniesing2006}. \citet{Vukadinovic2010} used a set of features for classifying the calcium candidate object of blood vessels. 
These features include smoothed intensity, Gaussian derivative features, and a set of shape features including spatial locations (distance to the lumen). 
\citet{Bogunovic2012a} extracted a set of labeled bifurcation feature vectors of vessels to train the classifier. 
\citet{Mehmet2016} extracted local features based on image intensity, intensity gradient, sample positions, and angles. 
They claimed that the Hessian-matrix-based features can aid in distinguishing between tubular and non-tubular structures.

One application of hand-crafted features is in the development of learning-based kernels.
\citet{Poletti2014c} learned a set of optimal discriminative convolution kernels to be used in AdaBoost classifications. 
The multi-kernel learning method proposed in \citep{Liu2014c} utilizes the features from the Hessian-matrix-based vesselness measures, multi-scale Gabor filter responses, and multi-scale line strengths. 
\citet{Lesage2016} learned the non-parametric kernel with likelihood terms of direction and radius transition priors. 
To estimate the vessel direction and diameter, \citet{Asl2017} formulated a kernelized covariance matrix from the training data.

Another application is the learning-based filters of vessels.
Assuming that the vessel properties changed constantly, \citet{Lin2012b} learned the continuity pattern of the current segment using the extended Kalman filter.
\citet{Azzopardi2013c} learned the appropriate prototype features in the filter-configuration process.
In addition to the appearance features of learning filters, \citet{Annunziata2015a} introduced the context information (i.e., relationships among objects) in the filter-learning process. 
The authors assumed that the learned context filters had two clear advantages: incorporating high-level information and obtaining high efficiency and adaptability.
To accelerate the learning process, \citet{Sironi2015} computed the filters by linearly combining them with a smaller number of separable filters. 
This operation can considerably address the problem of the computational complexity at no extra cost in terms of performance. 
\citet{Annunziata2016} proposed a warm-start strategy to solve this problem. 
This strategy is based on carefully designing hand-crafted filters and modeling appearance properties of curvilinear structures, which are then refined by convolutional sparse coding.
Using vascular filters (e.g., Frangi filter and optimally oriented flux), \citet{Zhang2017g} extracted the corresponding types of vascular features and integrated these feature responses into a structured random forest to classify voxels into positive and negative classes. \citet{Deng2018} named all the features for the random forest (RF) as discriminative integrated features. 
These features are classified as low-level features, vascular features, context features, and local self-similarity descriptor.
In addition to the kernels and filters, \citet{Javidi2017} constructed two separate dictionaries to learn the vessel and non-vessel representations. These learned dictionaries yield a strong representation containing the semantic concepts of the image.

\subsection{Classifications}\label{Classification}
The conventional machine-learning-based methods obtain vessels using classifiers.
For vessel-tracking tasks, the methodologies of classification can be broadly categorized into the unsupervised and supervised learning strategies.

Unsupervised learning-based methods train the classifier without using labeled vessel data or explicitly using any supervised classification techniques. 
To separate related regions, seeds and patches, the main schemes reported in the literature are clustering techniques (e.g., k-means and fuzzy C-means).
Instead of explicitly obtaining sparse representations, k-means clustering tends to discover sparse projections of the data \citep{Coates2012}. 
As a pre-processing step, the k-means algorithm can be used to partition the pixels into several clusters; e.g., three clusters of related regions \citep{Saffarzadeh2014} or five groups of images \citep{Zhang2014d}.
In the process, \citet{Jodas2017} employed the k-means algorithm with subtractive clustering to separate the vessel regions in the image according to the gray-scale intensity. 
The k-means algorithm is also used for the final refinement of vessel segmentation \citep{Goceri2017a}. 
In addition to vessel regions, \citet{Lu2017a} used the manually annotated seeds to represent the vascular features and utilized k-means clustering to exclude the wrong seeds. 
To find the representative patches from numerous candidate patches, \citet{Xia2018} used k-means clustering to group the patches under the Euclidean distance metric.
Fuzzy C-means clustering is another unsupervised learning method for pattern recognition that employs various image properties for separation.
Image pixel intensities are not mutually independent; hense, \citet{Kande2010} used a thresholding technique based on the spatially weighted fuzzy C-means algorithm, which can well preserve the spatial structures in a binarized/thresholded image.
In \citep{Mapayi2015}, phase-congruency \citep{Kovesi1999} has been used to preserve the features with in-phase frequency components, and fuzzy C-means clustering is performed for accurate retinal vessel segmentation.
\citet{Mapayi2016} further investigated the difference image with fuzzy C-means for the detection of vessels in the retinal image.
An improved fuzzy C-means clustering in \citep{Khan2016} was also used for pixel classification based on the texture features.
Using the contrast-time curve of the pixel as inputs, \citet{Haddad2018} separated the major vessels from the capillary blush and background noise through fuzzy C-means clustering. 
\citet{Zeng2018b} constructed the intensity model based on kernel fuzzy C-means to extract the intensity feature of thick vessels.

The ground truth is absent; hense, the performance of unsupervised methods relies on particular features based on the statistical distribution of the overall input data. 
In contrast, supervised learning methods require a manually annotated set of training images for classifications. 
The extracted features and ground truth of every sample are collected to train the classifier.
Most of these methods in the supervised category use various classifiers (\Zxhreffig{fig: Traditional supervised learning})---support vector machine (SVM), boosting-based methods, and random forests---to distinguish the vascular patterns in the images.

The SVM classifier performs vessel or non-vessel classification by constructing an N-dimensional hyperplane that optimally separates the vessel samples into different categories. 
The classification ability of the SVM is based on the feature vectors obtained by different operators and vascular-dedicated filters.
The operators can be line operator \citep{Ricci2007}, Gabor filters \citep{A.Osareh2009}, and wavelets \citep{You2011}.
The vascular-dedicated filters can be Frangi filter \citep{Frangi1998} and optimally oriented flux \citep{Law2010}.
To distinguish between the vessels,  the feature vectors can also be formulated via general measures; e.g., distance between adjacent nodes \citep{Hanaoka2015}, geometric shapes \citep{Kang2015a}, and normal cross-sections \citep{Jawaid2017b}. 
To deal with more complex cases in the curve detection, \citet{Chen2015c} utilized the joint feature representations---e.g., smoothness of points and position relationships---for classification.
Instead of classifying the pixels, in the framework of the fully connected conditional random field (CRF), the SVM methods are employed to adjust the weight parameters in the energy function \citep{Orlando2017}.
\citet{Lorza2018} used the SVM with a radial basis function kernel to obtain the probability map of the vessel region.

%【boosting-based methods】
Boosting-based methods are dependent on building strong classification models from a linear combination of weak classifiers, making the training easier and faster. 
Simple functions, such as regression stumps, can be employed in boosting-based methods to detect curvilinear objects \citep{Turetken2016}. 
In vesse-segmentation tasks, a regression stump is a decision tree with two terminal nodes. 
For a given vascular feature, the tree selects a branch according to the threshold based on a binary decision function.
As a special case of boosting-based methods, the AdaBoost learning model is trained to automatically detect the bifurcation points of vessels with elaborately selected features. 
To improve the accuracy of classifications, numerous filters are necessary to obtain the vascular features \citep{Zhou2007b}. \citet{Lupascu2010} used feature vectors, which are composed of eight elements including the output of filters, measures, and other transformation results, to encode vascular information on the local intensity structure, spatial properties, and geometry at multiple scales.
\citet{Gu2017} constructed boosting-tree methods based on features such as variable sizes and locations. 
These features represent the encoded global contextual information (e.g., context distance and local spatial label patterns).
\citet{Memari2017} completed the segmentation based on feature extraction and selection steps, along with the AdaBoost classifier.
If a classification tree has an AdaBoost classifier at each node, then the tree will be the probabilistic boosting-tree (PBT) classifier. \citet{Zheng2011c} exploited the PBT to identify the pixel inside the vessel using 24 feature vectors from the Hessian matrix.
To improve the performance of retinal segmentation methods in the presence of lesions, an ensemble classifier of boosted \citep{Freund1995} and bagged decision trees \citep{Breiman1996} has been proposed to manage the healthy and pathological retinal images via several encoded features \citep{Fraz2012b}. 
The ensembles of bagged decision trees have also been employed to learn the mapping between vessel widths and corresponding points \citep{Lupascu2013e}. 
In \citep{Hashemzadeh2019}, a root-guided decision tree was used to distinguish between the vessel and non-vessel regions.

A collection of tree-structured classifiers can be assembled as an RF classifier.  
Different from the SVM and decision trees, the RF \citep{Cutler2012, Zhang2016g} tends to deliver high performance because of the embedded feature selection in the model-generation process.
In the vessel-tracking process, the selected features are invariably related to the intensity profile or vascular shapes (e.g., tubular structures \citep{Annunziata2015, Melki2014, Zhang2017d}, vessel center \citep{Schneider2015a}, and tree-like structures \citep{Sankaran2016}). 
To cover more features for the RF classifier, researchers used multiple techniques to generate representations in various spaces. \citet{Cherry2015} used two sets of features to distinguish the abnormal vessels: vessel cues and local information. 

To improve the performance and avoid the over-fitting problems, hybrid classifiers are used for vessel-classification problems. 
\citet{Rani2016} combined the SVM and tree-bagger techniques to distinguish between the vessel and non-vessel structures. 
The works of \citet{Lugauer2014a} and \citet{Chapman2015} used the RF, PBT, and logistic regression classifier to identify the lumen contours and edges of vessels. 
\citet{Hu2018e} intricately applied the cascade-AdaBoost-SVM classifiers to delineate the vessel boundaries.

\subsection{Statistical models}\label{Statistic models}
The profiles of intensity and geometries of vessels can be learned using statistical models.
\citet{Vukadinovic2010} determined the vessel calcium object threshold by simply observing the calcium objects disappear in the image dataset; this threshold is used to segment the calcium regions of the vessels. 
In \citep{Cheng2012a}, the vessel center threshold, which is referred to as the strong peak near the center of the profile, was learned from a set of manually-labeled samples of parallel linear structures.
For more detailed information, \citet{Schaap2011} learned the local point distribution models and a nonlinear boundary intensity model by statistically analyzing the set of annotated training data. 
\citet{Zheng2012a} divided the vessel model into four structures, each of which is recognized via learned detectors. Instead of designing individual classifiers for each geometrical constraint, \citet{Rempfler2014a} simply learned the global statistic of the desired geometrical properties of the network from the dataset.

Based on the probabilistic model, the topological structures of vessels, e.g., branches and connections, can also be learned for vessel tracking.
\citet{Rempfler2015b} learned the physiological geometric properties of vessels such as the relative frequencies of radii and deviation angles of vessel segments. 
\citet{Asl2017} learned these relationships using kernels. 
\citet{Zhao2017d} learned the topological tree and geometrical statistics of parameters including tree hierarchy, branch angle, and length statistics.
Contrary to the pixel-based and object-based methods, \citet{Chai2013} modeled the vessel connections using graph theory.
To describe the shape of the graph, they constructed three sets of parameters: graph connectivity, edge orientation, and line width,
These parameters can be learned from annotated image samples through a maximum likelihood estimation.
Considering the intensity distributions, \citet{Kalaie2017b} developed a directed probabilistic graphical model whose hyperparameters are estimated using a maximum likelihood solution based on Laplace approximation.

\vspace{-1em}
\section{Vessel tracking based on deep learning}\label{Vessel tracking based on deep learning}
\begin{sidewaystable*}
\begin{minipage}[]{1.1\textwidth}
\caption{Overview of deep-learning-based methods for tracking \textbf{retinal vessels}: the evaluation metrics and datasets are presented in Section \ref{Evaluation issues}; see list of abbreviations at the bottom.}
\label{tab:deep learning:retinal vessel}
\vspace{-0.2cm}
\centering
\scriptsize
\begin{threeparttable}
\begin{tabular}{p{2.8cm}p{2.6cm}p{3.8cm}p{5.4cm}p{6cm}}
\hline\hline
\multirow{1}{*}{Authors} &\multirow{1}{*}{Methods}&\multirow{1}{*}{Data}&\multirow{1}{*}{Experiment}&\multirow{1}{*}{Results}\\
\cmidrule(r){1-5}
\multirow{3}{*}{\citet{Li2016b}}	&\multirow{3}{*}{Encoder-decoder, 2D}	&Retinal colored image, DRIVE&20 for training, 20 for testing , one-off train + test&Acc=0.9527, AUC=0.9738, Se=0.7569, Sp=0.9816\\
{}&{}&Retinal colored image, STARE&19 for training, 1 for testing, leave-one-out&Acc=0.9628, AUC=0.9879, Se=0.7726, Sp=0.9844\\
{}&{}&Retinal colored image, CHASE-DB-1 &20 for training, 8 for testing, one-off train + test&Acc=0.9581, AUC=0.9716, Se=0.7507, Sp=0.9793\\
\cmidrule(r){1-5}
\multirow{2}{*}{\citet{Liskowski2016}}	&\multirow{2}{*}{CNN without pooling, 2D}	&Retinal colored image, DRIVE&20 for training, 20 for testing, one-off train + test&Acc=0.9495, AUC=0.9720\\
{}&{}&Retinal colored image, STARE&19 for training, 1 for testing, leave-one-out&Acc=0.9566, AUC=0.9785\\
\cmidrule(r){1-5}
\citet{Lahiri2017}	&GAN, 2D &Retinal colored image, DRIVE&20 for training, 20 for testing, one-off train + test&AUC=0.962\\
\multirow{1}{*}{\citet{Costa2018}}	&\multirow{1}{*}{GANs, 2D}	&Retinal colored image, DRIVE&20 for training, 20 for testing, one-off train + test&AUC=0.841\\
\cmidrule(r){1-5}
\multirow{2}{*}{\citet{Guo2018b}}	&\multirow{2}{*}{Multiple CNNs, 2D}	&Retinal colored image, DRIVE&20 for training, 20 for testing, one-off train + test&Acc=0.9597, AUC=0.9726\\
{}&{}&Retinal colored image, STARE&20 for testing&Acc=0.9613, AUC=0.9737\\
\cmidrule(r){1-5}
\multirow{3}{*}{\citet{Yan2018e}}	&\multirow{3}{*}{Encoder-decoder, 2D}	&Retinal colored image, DRIVE&20 for training, 20 for testing, one-off train + test&Acc=0.9542, AUC=0.9752, Se=0.7653, Sp=0.9818\\
{}&{}&Retinal colored image, STARE&19 for training, 1 for testing, leave-one-out&Acc=0.9612, AUC=0.9801, Se=0.7581, Sp=0.9846\\
{}&{}&Retinal colored image, CHASE-DB1&20 for training, 8 for testing, one-off train + test&Acc=0.9610, AUC=0.9781, Se=0.7633, Sp=0.9809\\
{}&{}&Retinal colored image, HRF&5 for training, 40 for testing, one-off train + test&Acc=0.9437, Pr=0.6647, Se=0.7881, Sp=0.9592\\
\cmidrule(r){1-5}
\multirow{2}{*}{\citet{Wu2018b}}	&\multirow{2}{*}{Multiple CNNs, 2D}	&Retinal colored image, DRIVE&20 for training, 20 for testing, one-off train + test&Acc=0.9567, AUC=0.9807, Se=0.7844, Sp=0.9819\\
{}&{}&Retinal colored image, CHASE-DB1&20 for training, 8 for testing,  one-off train + test&Acc=0.9637, AUC=0.9825, Se=0.7538, Sp=0.9847\\
\cmidrule(r){1-5}
\multirow{3}{*}{\citet{Zhao2018f}}	&\multirow{3}{*}{GANs, 2D}	&Retinal colored image, DRIVE&20 for training, 20 for testing, one-off train + test&Se=0.8038, Sp=0.9815\\
{}&{}&Retinal colored image, STARE&10 for training, 10 for testing, one-off train + test&Se=0.7896, Sp=0.9841\\
{}&{}&Retinal colored image, HRF&22 for training, 23 for testing, one-off train + test&Se=0.8001, Sp=0.9823\\
\cmidrule(r){1-5}
\multirow{3}{*}{\citet{Zhang2018g}}	&\multirow{3}{*}{U-net, 2D}	&Retinal colored image, DRIVE&20 for training, 20 for testing, one-off train + test&Acc=0.9504, AUC=0.9799, Se=0.8723, Sp=0.9618\\
{}&{}&Retinal colored image, STARE&15 for training, 5 for testing, four-fold cross-validation&Acc=0.9712, AUC=0.9882, Se=0.7673, Sp=0.9901\\
{}&{}&Retinal colored image, CHASE-DB1&21 for training, 7 for testing, four-fold cross-validation&Acc=0.9770, AUC=0.9900, Se=0.7670, Sp=0.9909\\
\cmidrule(r){1-5}
\multirow{1}{*}{\citet{Gu2019a}}	&\multirow{1}{*}{Context encoder network}	&\multirow{1}{*}{Retinal colored image, DRIVE}&\multirow{1}{*}{20 for training, 20 for testing, one-off train + test}&\multirow{1}{*}{Acc=0.955, AUC=0.978}\\
\cmidrule(r){1-5}
\multirow{4}{*}{\citet{Jin2019}}	&\multirow{4}{*}{Deformable U-net, 2D}	&Retinal colored image, DRIVE&20 for training, 20 for testing, one-off train + test&Acc=0.9566, AUC=0.9802, TNR=0.9800, TPR=0.7963\\
{}&{}&Retinal colored image, STARE&19 for training, 1 for testing, leave-one-out&Acc=0.9641, AUC=0.9832, TPR=0.7595, TNR=0.9878\\
{}&{}&Retinal colored image, CHASE-DB1 &14 for training, 14 for testing, one-off train + test&Acc=0.9610, AUC=0.9804, TNR=0.9752, TPR=0.8155\\
{}&{}&Retinal colored image, HRF & 15 for training, 30 for testing, one-off train + test&Acc=0.9651, AUC=0.9831, TNR=0.9874, TPR=0.7464\\	
\cmidrule(r){1-5}
\multirow{2}{*}{\citet{Lian2019}}	&\multirow{1}{*}{U-net, Res-net and }	&Retinal colored image, DRIVE&20 for training, 20 for testing, one-off train + test&Acc=0.9692, Pr=0.8637, Se=0.8278, Sp=0.9861\\
{}&{attention scheme, 2D}&Retinal colored image, STARE&10 for training, 10 for testing, one-off train + test&Acc=0.9740, Pr=0.8823, Se=0.8342, Sp=0.9916\\
\cmidrule(r){1-5}
\multirow{3}{*}{\citet{Mou2019}}	&\multirow{3}{*}{Dense dilate network, 2D}	&Retinal colored image, DRIVE&20 for training, 20 for testing, one-off train + test&Acc=0.9594, AUC=0.9796, Se=0.8126, Sp=0.9788\\
{}&{}&Retinal colored image, STARE&15 for training, 5 for testing, four-fold cross-validation&Acc=0.9685, AUC=0.9858, Se=0.8391, Sp=0.9769\\
{}&{}&Retinal colored image, CHASE-DB1&21 for training, 7 for testing, four-fold cross-validation&Acc=0.9637, AUC=0.9812, Se=0.8268, Sp=0.9773\\
\cmidrule(r){1-5}
\multirow{4}{*}{\citet{Shin2019}}	&\multirow{4}{*}{CNN + GNN, 2D}	&Retinal colored image, DRIVE&20 for training, 20 for testing , one-off train + test&Acc=0.9271, AUC=0.9802, Se=0.9382, Sp=0.9255\\
{}&{}&Retinal colored image, STARE&10 for training, 10 for testing, one-off train + test&Acc=0.9378, AUC=0.9877, Se=0.9598, Sp=0.9352\\
{}&{}&Retinal colored image, CHASE-DB1 &20 for training, 8 for testing, one-off train + test&Acc=0.9373, AUC=0.9830, Se=0.9463, Sp=0.9364\\
{}&{}&Retinal colored image, HRF & 15 for training, 30 for testing, one-off train + test&Acc=0.9349, AUC=0.9838, Se=0.9546, Sp=0.9329\\
\cmidrule(r){1-5}
\multirow{3}{*}{\citet{Cherukuri2020}}	&\multirow{3}{*}{CNN + geometric prior, 2D}	&Retinal colored image, DRIVE&20 for training, 20 for testing, four-fold cross-validation&Acc=0.9563, AUC=0.9814\\
{}&{}&Retinal colored image, STARE&10 for training, 10 for testing, one-off train + test&Acc=0.9687, AUC=0.9903\\
{}&{}&Retinal colored image, CHASE-DB1 &14 for training, 14 for testing, one-off train + test&Acc=0.9672, AUC=0.9833\\
\cmidrule(r){1-5}
\multirow{2}{*}{\citet{Ding2020}}	&\multirow{1}{*}{Pre-trained model, vessel}	&\multirow{2}{*}{Retinal UWF FP, PRIME-FP20}&\multirow{2}{*}{15 images, four-fold cross-validation}&\multirow{2}{*}{AUCPR=0.842, Max DSC=0.772}\\
{}&{maps, and noise label, 2D}&{}& {}&{}\\
\hline
\hline
\end{tabular}
\begin{tablenotes}
        \footnotesize
        \item[*] List of abbreviations: Acc=accuracy; AUC=area under the ROC curve; CNN=convolutional neural networks; GAN=generative adversarial network; GNN=graph neural network; Pr=precision; Re=recall;  Se=sensitivity; Sp=specificity; TNR=true negative rate; TPR=true positive rate; UWF FP=ultra-widefield fundus photography .
      \end{tablenotes}
\end{threeparttable}

\end{minipage}
\end{sidewaystable*}
\begin{sidewaystable*}
\begin{minipage}[]{1.1\textwidth}
\caption{Overview of deep-learning-based methods for tracking \textbf{coronary vessels}: the evaluation metrics and datasets are presented in Section \ref{Evaluation issues}; see list of abbreviations at the bottom.}
\label{tab:deep learning:coronary}\vspace{-0.2cm}
\centering
\scriptsize
\begin{threeparttable}
\begin{tabular}{p{2.2cm}p{2.5cm}p{4.5cm}p{5.5cm}p{5.2cm}}
\hline\hline
\multirow{1}{*}{Authors} &\multirow{1}{*}{Methods}&\multirow{1}{*}{Data}&\multirow{1}{*}{Experiment}&\multirow{1}{*}{Results}\\
\cmidrule(r){1-5}
\citet{Lee2019b}	&CNN + Shape prior, 3D &Coronary CTA, local data&274 for training, 136 for testing, one-off train + test&DSC=0.768, HD=3.55mm\\
\cmidrule(r){1-5}
\multirow{1}{*}{\citet{Shin2019}}	&\multirow{1}{*}{CNN + GNN, 2D}	&Coronary x-ray, CA-XRA (local data) & 2958 for training, 179 for testing, one-off train + test&Acc=0.9517, AUC=0.9914, Se=0.9700, Sp=0.9507\\		
\cmidrule(r){1-5}
\multirow{3}{*}{\citet{Wolterink2019}}	&\multirow{3}{*}{\tabincell{l}{CNN with dilated-\\convolution, 3D}}	&Coronary CTA, CAT08&7 for training, 1 for testing, leave-one-out&AI=0.21mm, OF=0.815, OT=0.970, OV=0.937\\
{}&{}&Coronary CTA, UMCU dataset (local data)&8 for training, 50 for testing, one-off train + test&Median of radius=0.81mm\\
{}&{}&Coronary CTA, orCaScore&8 for training, 36 for testing, one-off train + test&Visual check\\
\hline
\hline
\end{tabular}
\begin{tablenotes}
        \footnotesize
        \item[*] List of abbreviations: 
        \item[*] Acc=accuracy; AI=average inside; AUC=area under the ROC curve; CNN=convolutional neural networks; CTA = computed tomography angiography; GNN=graph neural network; DSC=dice similarity coefficient; HD=hausdorff distance; Se=sensitivity; Sp=specificity; OF=overlap until first error; OT=overlap with the clinically relevant part of the vessel; OV=overlap.
      \end{tablenotes}
\end{threeparttable}

\end{minipage}\vspace{0.1cm}
\begin{minipage}[]{1.1\textwidth}
\caption{Overview of deep-learning-based methods for tracking \textbf{other vessels}: the evaluation metrics and datasets are presented in Section \ref{Evaluation issues}; see list of abbreviations at the bottom.}
\label{tab:deep learning:other vessels}\vspace{-0.2cm}
\centering
\scriptsize
\begin{threeparttable}
\begin{tabular}{p{2.8cm}p{2.6cm}p{4.3cm}p{5.2cm}p{3.8cm}}
\hline\hline
\multirow{1}{*}{Authors} &\multirow{1}{*}{Methods}&\multirow{1}{*}{Data}&\multirow{1}{*}{Experiment}&\multirow{1}{*}{Results}\\
\cmidrule(r){1-5}
\citet{Marques2016}	&U-net, 3D &Multiple organs, CT, MR, local data&67 for training, 19 for testing, one-off train + test&Pr=0.362\\
\cmidrule(r){1-5}
\multirow{3}{*}{\citet{Huang2018}}	&\multirow{3}{*}{Unet, 3D}	&Liver contrast-enhanced CT, 3D-IRCADb&10 for training, 10 for testing, one-off train + test&DSC=0.675, Se=0.743\\
{}&{}&Liver CT, SLIVER07&20 for testing&Visual check\\
{}&{}&Liver CT, local data&10 for testing&Visual check\\
\cmidrule(r){1-5}
\multirow{1}{*}{\citet{Lian2018a}}	&\multirow{1}{*}{Modified U-net, 2D}&Perivascular spaces, 7T MR, local data&6 for training, 14 for testing, one-off train + test&DSC=0.77, PPV=0.83, Se=0.74\\
\cmidrule(r){1-5}
\citet{Nardelli2018}	&CNN + graph cut, 3D &Pulmonary CT, local data&4 for training, 16 for validation&Acc=0.87\\
\cmidrule(r){1-5}
\multirow{2}{*}{\citet{Kitrungrotsakul2019}}	&\multirow{2}{*}{DensNet, 2.5D}	&hepatic MR, IRCAD&19 for training, 1 for testing, leave-one-out&DSC=0.903, Se=0.929\\
{}&{}&hepatic MR, VASCUSYNTH&9 for training, 1 for testing,  leave-one-out&DSC=0.901\\
\cmidrule(r){1-5}
\multirow{2}{*}{\citet{He2020a}}	&Auto-encoder + &\multirow{2}{*}{Renal artery, CT, local data}&\multirow{2}{*}{52 for training, 104 for testing, one-off train + test}&\multirow{2}{*}{DSC=0.884, HD=25.439mm}\\
{}&{dense bias connection, 3D}&{}&{}&{}\\
\cmidrule(r){1-5}
\multirow{2}{*}{\citet{Nazir2020}}	&CNN + dilated-&\multirow{2}{*}{Intracranial vessel, CTA, local data}&\multirow{2}{*}{50 for training, 20 for testing, one-off train + test}&\multirow{2}{*}{DSC=0.8946, HD=5.04mm}\\
{}&{convolution, 3D}&{}&{}&{}\\
\cmidrule(r){1-5}
\multirow{1}{*}{\citet{Ni2020}}	&CNN + channel attention&\multirow{1}{*}{Intracranial vessel, CTA, local data}&\multirow{1}{*}{9488 slices for training, 480 images for testing}&\multirow{1}{*}{DSC=0.965}\\
\hline
\hline
\end{tabular}
\begin{tablenotes}
        \footnotesize
        \item[*] List of abbreviations: Acc=accuracy; CNN=convolutional neural networks; CT=computed tomography; CTA = computed tomography angiography; DSC= dice similarity coefficient; HD=hausdorff distance; IRCAD: \url{http://www.ircad.fr}; MR=magnetic resonance; Pr=precision; PPV=positive predictive value; Se=sensitivity; VASCUSYNTH \citep{2010VascuSynth}.
      \end{tablenotes}
\end{threeparttable}

\end{minipage}
\end{sidewaystable*}

\begin{sidewaystable*}
\vspace{0.8cm}
 \begin{minipage}[t]{0.6\textwidth}\vspace{-0.8cm}
  \centering
\caption{Overview of techniques in deep-learning-based methods: see list of abbreviations at the bottom.}
\label{tab:deep learning techniques}\vspace{-0.2cm}
\begin{threeparttable}
           \begin{tabular}{p{5cm}p{4.5cm}p{13cm}}
\hline\hline
\multirow{1}{*}{Deep-learning-based methods} &\multirow{1}{*}{Techniques}&\multirow{1}{*}{}\\
\cmidrule(r){1-3}
\multirow{24}{*}{Network architectures}&\multirow{1}{*}{CNN + Dilated convolution:}&\citet{Wolterink2019, Mou2019, Nazir2020}\\ 
\cmidrule(r){2-3}
{}&\multirow{1}{*}{CNN + Deformation convolution:} &\citet{Jin2019}\\
\cmidrule(r){2-3}
{}&\multirow{1}{*}{CNN + No Pooling:} &\citet{Liskowski2016, Tetteh2017a}\\
\cmidrule(r){2-3}
{}&\multirow{2}{*}{CNN + Probability maps:} &\citet{Ganin2015, Fu2016b, Mo2017}\\
{}&{}&\citet{Hu2018, Uslu2019, Lin2019, Ding2020}\\
\cmidrule(r){2-3}
{}&\multirow{1}{*}{CNN + Attention mechanism:} &\citet{Shen2019, Li2019, Lian2019, Ni2020}\\
\cmidrule(r){2-3}
{}&\multirow{1}{*}{CNN + Skipping/short connection:} &\citet{Feng2018, Guo2019}\\
\cmidrule(r){2-3}
{}&\multirow{1}{*}{CNN + CRF:} &\citet{Fu2016a, Luo2017}\\
\cmidrule(r){2-3}
{}&\multirow{1}{*}{CNN + Prior:} &\citet{Lee2019b, Cherukuri2020}\\
\cmidrule(r){2-3}
{}&\multirow{1}{*}{Multi-task learning:} &\citet{Maninis2016, Tan2017}\\
\cmidrule(r){2-3}
{}&\multirow{2}{*}{Encoder-decoder:} &\citet{Li2016b, Fan2017, Dasgupta2017}\\
{}&{}&\citet{Gu2019a, He2020a}\\
\cmidrule(r){2-3}
{}&\multirow{1}{*}{U-net:} &\citet{Fan2018, Huang2018}\\
\cmidrule(r){2-3}
{}&\multirow{2}{*}{Modified U-net:} &\citet{Chen2018j, Kandil2018a, Wang2019c, Zhang2019}\\
{}&{}&\citet{Girard2019, Zhang2018g, Dharmawan2019, Zhang2019}\\
\cmidrule(r){2-3}
{}&\multirow{1}{*}{GNN:} &\citet{Shin2019}\\
\cmidrule(r){2-3}
{}&\multirow{1}{*}{GANs:} &\citet{Costa2018, Yu2019}\\
\cmidrule(r){2-3}
{}&\multirow{1}{*}{Multiple CNNs:} &\citet{Wu2018b, Guo2018b}\\
\cmidrule(r){1-3}
\multirow{5}{*}{Pre-processing:}&\multirow{1}{*}{Contrast/brightness normalization: }&\citet{Vega2015b, Liskowski2016}\\ 
\cmidrule(r){2-3}
{}&\multirow{1}{*}{Whiting:} &\citet{Liskowski2016, Marques2016}\\
\cmidrule(r){2-3} 
{}&\multirow{2}{*}{Augmentation:} &\citet{Huang2018, Guo2019, Fan2018}\\ 
{}&{}&\citet{Livne2019, Zreik2018a, Lin2019}\\
\cmidrule(r){1-3} 
\multirow{5}{*}{Sampling strategies}&\multirow{1}{*}{Patch as samples:}&\citet{Nardelli2018a, Wolterink2019}\\ 
\cmidrule(r){2-3}
{}&\multirow{1}{*}{Image as samples;} &\citet{Hu2018, Mo2017}\\
\cmidrule(r){2-3} 
{}&\multirow{2}{*}{Processed image as samples:} &\citet{Nardelli2017, Nardelli2018, Hajabdollahi2018}\\
{}&{}&\citet{Zhao2018g, Zreik2018a}\\
\cmidrule(r){1-3} 
\multirow{7}{*}{Loss functions:}&\multirow{2}{*}{Loss based on cross-entropy:}&\citet{Dasgupta2017, Nardelli2018a, Wu2018b, Jin2019}\\ 
{}&{}&\citet{Dharmawan2019, Mo2017,Guo2019, Lin2019}\\
\cmidrule(r){2-3}
{}&\multirow{2}{*}{Loss to tackle data imbalance:} &\citet{Li2018b, Zhang2018g, Hu2018, Livne2019}\\
{}&{}&\citet{Soomro2019b, Kitrungrotsakul2019, Huang2018, Lian2018a}\\
\cmidrule(r){2-3} 
{}&\multirow{1}{*}{Loss based on squared error:} &\citet{Li2016b, Fan2017}\\\cmidrule(r){2-3} 
{}&\multirow{1}{*}{More complex:} &\citet{Yan2018e, Jiang2019}\\
\hline
\hline
\end{tabular}
\end{threeparttable}
\begin{tablenotes}
        \footnotesize 
        \item[*] List of abbreviations: CNN=convolutional neural networks; CRF=conditional random field; GAN=generative adversarial network; GNN=graph neural network.
      \end{tablenotes}
 \end{minipage}
    \end{sidewaystable*}

Using deep-learning-based methods, deep neural networks can be developed to map the input data into vascular patterns such as center points and vascular regions. 
These patterns can be used to obtain the vessels directly or indirectly. 
To this end, various deep-learning techniques have been proposed.
This section reviews the deep-learning-based methods from three aspects: frameworks of vessel tracking (Section \ref{Frameworks of vessel tracking}), architecture of deep neural networks (Section \ref{Architecture of networks}), and model training (Section \ref{Training of CNN models}).
\Zxhreftbs{tab:deep learning:retinal vessel} - \ref{tab:deep learning:other vessels}  summarize the decompositions of a selection of works which are representative of the main trends in the field according to the applications.
\Zxhreftbs{tab:deep learning techniques} summarizes the existing deep-learning-based works by grouping them into different subcategories.
\subsection{Frameworks of vessel tracking}\label{Frameworks of vessel tracking}

Vessel tracking can be achieved using hierarchical features via a unified framework or a two-step processing scheme. 
The unified vessel-tracking methods are implemented by integrating feature extraction and pixel classification into one network. 
In contrast, the two-step scheme generally employs a conventional method to track the vessel based on the preceding vessel features extracted using a deep convolutional neural network (CNN).

The unified framework of vessel tracking can be transformed into resolving a classification or regression problem via the fully connected layers of CNN. 
The output neurons of CNN that are generally connected to the fully connected layers of the network determine the labels of pixels. 
To separate vascular regions from the background using CNN, two neurons are typically set as output layers, following the fully connected layers \citep{Liskowski2016, Marques2016, Dasgupta2017, Hu2017a, Oliveira2017}. 
More neurons are output simultaneously to segment vessels and other structures \citep{Maninis2016, Tan2017}; this can be regarded as multi-task learning.
The idea of multiple tasks has been extended in \citep{Lahiri2017}, where a discriminator-classifier network differentiates between fake and real vessel samples and assigns correct class labels.
The neurons of fully connected layers in conventional CNNs have large receptive fields of input; hence, the results are extremely coarse in the pixel-level vessel segmentation. 
To resolve the problem, \citet{Li2016b} improved the coarse results of conventional CNNs by outputting label maps of the same size. 
A pixel in the label maps can be affected by the multiple image patches in its neighborhood. 
This idea is similar to the fully connected CRF, which considers the relationships among the neighbor pixels.
To achieve vessel segmentation, the CRF layers can also be used after using the convolutional layers \citep{Fu2016a, Luo2017}.

The vessel-tracking process can be divided into two steps: feature learning and vessel tracking.
In feature learning, the CNN maps the input image into intermediate representations located between the input and tracking results; e.g., probability maps \citep{Khowaja2016, Wu2016a, Nasr-Esfahani2016, Wolterink2019, Mou2019}, geometric priors \citep{Cherukuri2020}, and other feature maps \citep{Wang2015}.
In vessel tracking, the conventional tracking method can be applied to these intermediate representations. 
The simple approach to complete the tracking is thresholding the probability map \citep{Nasr-Esfahani2016, Mo2017, Nasr-Esfahani2018}. 
\citet{Wang2015} employed ensemble RFs to classify the vascular pixels based on the output feature maps from the selected layers of CNN. 
\citet{Guo2018b} used a voting scheme to determine the results obtained by the CNN. 
\citet{Mou2019} performed vessel tracking by integrating the predicted probability maps and local vessel directions into the regularized walk algorithm. 
To refine the results of CNNs, \citet{Hu2018} added CRF modules at the end of the network and \citet{Chu2013} used the rank-1 tensor-approximation approach to complete the tracking.
Inspired by the label-propagation steps of registration methods, \citet{Lee2019b} employed a CNN to learn the deformations between the source and the target vessels.
The authors assumed that this template transformer network can provide guarantees on the resulting shapes of vessels.

\subsection{Network architectures}\label{Architecture of networks}
In vessel-tracking tasks, the CNNs are widely adopted for identifying hierarchical vascular features. 
To design an effective CNN for the recognition of vascular patterns, two aspects require thorough investigation: network components and integration of multiple networks.

\subsubsection{Network components}\label{Components of the networks}

\begin{figure*}[!t]
\setlength{\abovecaptionskip}{-0.cm}
\setlength{\belowcaptionskip}{-0.cm}
\centering
\includegraphics[scale=.7]{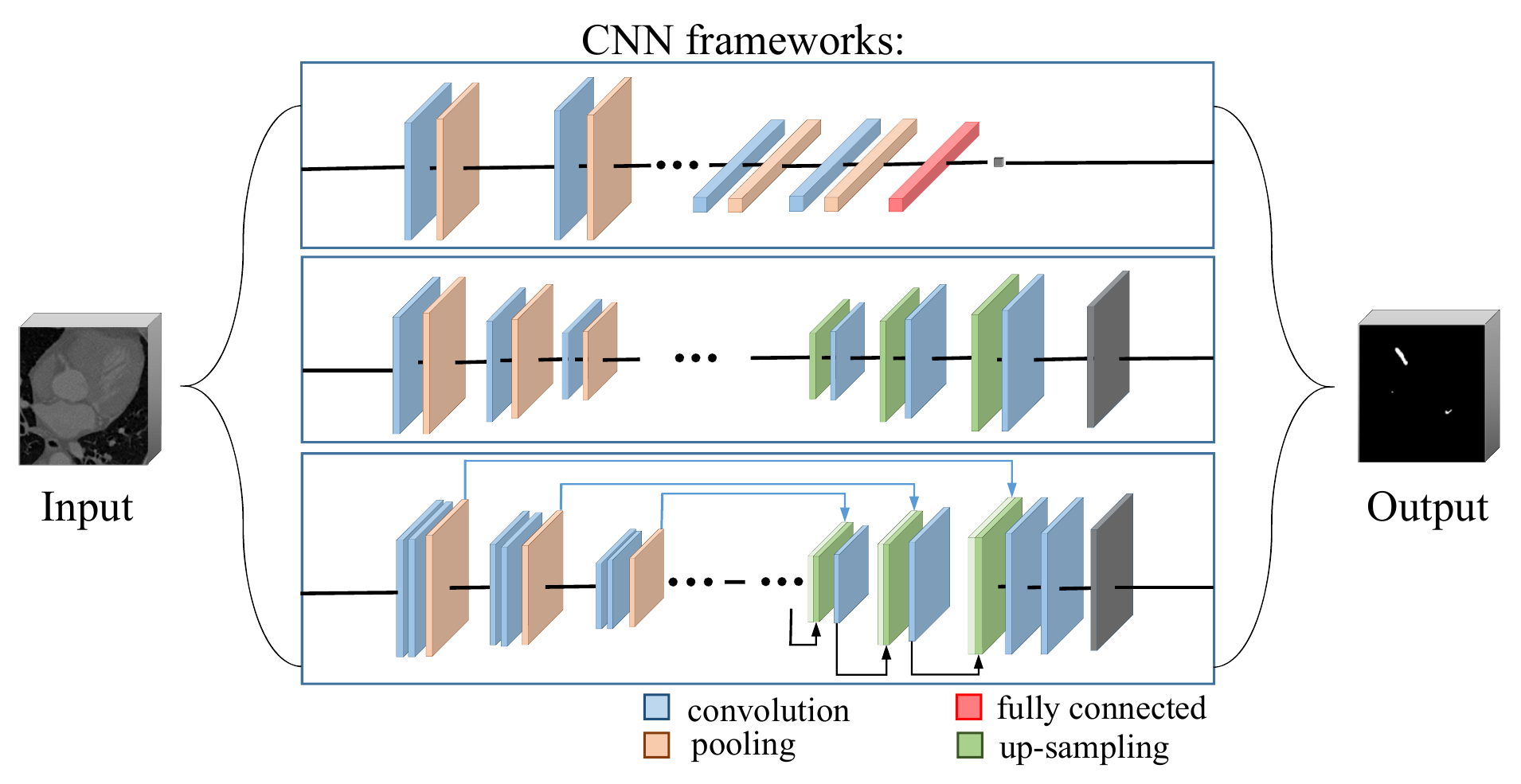}
\caption{Illustration of three selected CNN frameworks for the coronary segmentation: pixel-wise CNN (top), encoder-decoder (middle), and U-net (bottom).}\label{CNN frameworks}
\end{figure*}

A CNN is composed of a series of layers (\Zxhreffig{CNN frameworks}), typically including the convolutional layers, pooling layers, and fully connected layers. 
The convolutional and pooling layers are used to build the CNNs in the early applications of vascular feature extraction \citep{Chu2013}, whereas fully connected layers are usually added at the end of a CNN as  a part of pixel-classification tasks (Section \ref{Classification}). 
The convolutional layers activate the localized vascular features of the image and feature map by using a set of convolutional units \citep{Nardelli2018a, Zreik2018a}. 
A stack of dilated convolutions is used in convolutional layers \citep{Wolterink2019, Mou2019} to aggregate the features over multiple scales. 
In addition to dilated convolution modules, \citet{Nazir2020} adopted the inception module fusion of residual connection, enabling the network to capture advanced visual information under a controlled computational complexity.
To capture the various shapes and scales of vessels, \citet{Jin2019} integrated the deformation convolution into the network.
After the convolution layers, the pooling layers in the CNNs nonlinearly down-sample the input representation and preserve the feature information in each sub-region. 
The pooling layer aids in reducing the number of parameters irrelevant to the problem \citep{Nardelli2018a, Lian2018a, Zreik2018a}.
However, the scaling operation of the pooling layer is considered to rapidly reduce the already extremely limited information contained in the potentially small patch, causing the classification to be more exigent. 
Therefore, \citet{Liskowski2016} constructed a NO-POOL architecture, which performs well on the datasets. 
\citet{Tetteh2017a} also found that the pooling operations can lead to the loss of fine local details, which are extremely crucial in pixel-wise tasks. 
To solve this problem, they removed all the pooling layers of the CNN, making the feature-extraction layers robust enough to objects of interest of any size.
By employing various size-pooling operations, \citet{Gu2019a} used the residual multi-kernel pooling layer that encodes the multi-scale context features without extra learning weights.

Feature maps are organized sets of units obtained through convolution operations.
In vessel segmentation, different spatial forms of feature maps can be used in the CNNs.
Using a three-dimensional (3D) CNN, \citet{Jin2017} generated 3D feature maps to learn the structure variations in the 3D space. 
They assumed that 3D spatial information (especially the 3D branch-level continuity) and junction-bifurcating patterns are important for segmenting vascular structures.
Owing to high computational demands, they selected a relatively small region of interest (ROI) and trimmed the network. 
\citet{Yun2019} used a 2.5D CNN, which simultaneously takes three adjacent slices in each of the orthogonal directions, including axial, sagittal, and coronal, to improve the segmentation accuracy. 
However, they assumed that they could use the 3D CNN to entirely capture the 3D vascular information in its 3D convolutional layers.

The feature maps created in the network can be applied for final vessel-tracking tasks. 
\citet{Ganin2015} found that CNNs are insufficient for learning the mapping from the image patch for vessel annotation, leading to a severe under-fitting during the training and suboptimal performance during the test period. 
To resolve this issue, the network maps the input image or patches into intermediate representations using the CNN. 
In vessel-segmentation tasks, these mapping results may be the probability maps of vessels. 
By applying the sigmoid activation function to the final convolution layer, the CNN output is converted to the probability values in the foreground and background regions. 
The final predicted vessel segmentation can be obtained by fusing these probability maps, which describe the probability distributions of vessels and non-vessels. 
To predict the vessel boundary, \citet{Fu2016b} utilized the full CNN architecture to output the vessel probability map, followed by the CRF, for a binary segmentation. 
\citet{Mo2017} generated a weighted fusion layer by fusing the multi-scale feature maps from each branch output. 
In their framework, the probability map is computed using sigmoid functions on the fusion of feature maps. 
\citet{Hu2018} obtained the probability map using a multi-scale CNN.
By fusing the middle layer feature maps, this CNN model fuses richer multi-scale information for comprehensive feature description to learn more detailed information on the retinal vessels. 
\citet{Uslu2019} produced the probability maps of vessel interior, centerline, and edge locations. 
The authors assumed that the probability map can better explain the uncertainty and subjectivity of detecting vessels, especially those at the edge locations appearing in the ground truth data. 
To formulate the vessel segmentation as a style-transfer problem, \citet{Ding2020} used the binary probability maps as the tentative training data and the style targets.
Considering that shallower side-outputs capture rich detailed information, and deeper side-outputs have high-level but fuzzy knowledge,  \citet{Lin2019} outputted the feature maps of each intermediate layer using VGGNet \citep{Simonyan2014}.

\subsubsection{Integration of multiple networks}\label{Integration of multiple networks}

Because the designed forms of feature maps influence the performance of CNNs, feature maps of different layers are fused to further describe the vessels \citep{Wang2015, Fu2016b}. 
These feature-map forms can be derived by the various architectures of CNNs.
Based on the number of CNNs used in the vessel-segmentation task, the architecture can be designed as a single CNN or multiple CNNs.

A single CNN for vessel tracking can extract meaningful vascular-structures representations; this is regarded as a problem of dimension reduction or sparse coding feature spaces.
For this problem, encoder-decoder architectures (\Zxhreffig{CNN frameworks}) are introduced to encode the hierarchical features. 
Instead of transforming the input into another space, \citet{Li2016b} developed an auto-encoder network to learn the features, which could recover the input. 
The auto-encoder network can be embedded into the CNNs to extract features, which could manage large inter-anatomy variations and thin structures such as renal arteries \citep{He2020a}.
Inspired by the auto-encoder network, \citet{Fan2017} developed an encoder-decoder style network to learn the mapping functions from the images to the vessels.
By formulating the vessel-tracking problem as a multi-label inference problem, \citet{Dasgupta2017} used the encoder-decoder framework to learn the class-label dependencies of neighboring pixels. 
By employing the skip connections between the encoder and decoder layers, the modified auto-encoder network facilitates the proper memorization of global and local features and alleviates the vanishing gradient problem of deep CNNs. 
\citet{Feng2018} concatenated different feature maps through a skipping connection. 
To learn more inherent features from different scales, \citet{Guo2019} further exploited short connections to fuse multiple outputs of side output layers.
To fuse multi-scale features, \citet{He2020a} employed a dense biased connection that compresses and transmits the feature maps in each layer to every forward layer.
The authors assumed that this connection can reduce feature redundancy and maintain the integrity of the information and gradient flows.
\citet{Shin2019} used a graph neural network as a unified CNN to learn the vessel structures. 
%Considering both local appearances and global vessel structures, a graph neural network (GNN) into a unified CNN architecture to learn the graphical vessel structure \citep{Shin2019}.

Similar to an encoder-decoder architecture, the U-net \citep{ronneberger2015u} can extract vascular features using skipping connections (\Zxhreffig{CNN frameworks}). 
One feature map generated from a lower layer was concatenated to a corresponding higher layer. 
The U-net has been used to segment the coronary arteries in X-ray angiograms \citep{Fan2018} and liver vessels in CT images \citep{Huang2018}.
The global contextual information from the low-level features and the spatial details from the previous convolution guide the precise segmentation. 
Several methods attempt to efficiently extract or fuse vascular features by improving the structures of U-net \citep{Chen2018j, Kandil2018a, Wang2019c}. 
\citet{Yan2018e} added two separate branches at the end of U-net to simultaneously train the model with the segment-level and the pixel-wise losses. 
To improve the robustness and facilitate convergence, \citet{Zhang2018g} applied a residual connection inside each resample block, which added feature maps before the convolution layers. 
To reduce the over-fitting problems, the elements in the U-net framework were modified; e.g., adding a dropout layer \citep{Dharmawan2019} and reducing the number of channels \citep{Livne2019}. 
\citet{Zhang2019} modified the original U-net by applying an additional convolutional layer before implementing concatenation using the corresponding decoder layer. 
This configuration also aids in transferring low-dimensional features to a higher-dimensional space.

However, the general U-net may fail to extract some minuscule vessels because this feature accumulation is limited by the depth of U-net \citep{Jin2019}; accordingly, modified U-Nets are developed to focus on vascular structures. 
\citet{Jin2019} developed a deformable CNN to capture various vessel shapes and scales via deformable receptive fields, which are adaptive to input features. 
To highlight the vessel-like structures, the attention gate (AG) mechanism is introduced into the CNN \citep{Shen2019, Li2019}. 
This AG mechanism can highlight salient features and gradually suppress the characteristic response in unrelated background regions without passing multi-level information \citep{Shen2019, Li2019}. 
\citet{Lian2019} incorporated a weighted attention mechanism into the U-net framework.
Using this mechanism, the network only focuses on the target ROI and eliminates the irrelevant background noise.
To better learn the global feature information, \citet{Ni2020} introduced the channel attention mechanism when aggregating high-level and shallow features.

Different from the single CNN framework, multiple CNNs can be jointly adopted in a framework for vessel tracking. These CNNs can be designed according to different views of the image; e.g., three views (sagittal, coronal, and transverse) of patches \citep{Kitrungrotsakul2017a}. 
\citet{Guo2018b} employed multiple CNNs as a voted classifier to improve the performance. 
\citet{Zhao2018g} accomplished the voxel-level vessel segmentation via the hierarchical update of CNNs. 
The authors assumed that this network absorbed the learning experience of the previous iteration, which gradually transformed a semi-supervised task into a supervised one. 
\citet{Zhang2019} proposed a more complicated cascade U-net network (i.e., three sub-networks are designed for different detection tasks).

\subsection{Training strategies}\label{Training of CNN models}
The successful training of a useful CNN model relies on a series of strategies. 
It is essential to design a suitable training strategy to ensure that the network can focus on vascular regions.
Considering the vascular profiles and features, the training strategies should be carefully designed while considering the following aspects: pre-processing, sampling strategies, and formulation of loss functions.

\subsubsection{Pre-processing}\label{Pre-processing}
The trained networks tend to perform better on appropriately pre-processed images. The data pre-processing techniques include contrast/brightness normalization, whitening, and augmentation.

The image brightness may vary across the fields of view, affectings the network performance. 
To resolve this problem, the contrast or brightness normalization abstracts from these fluctuations and further focuses on vessel regions. 
A Gaussian kernel is used to homogenize the background \citep{Vega2015b}. 
\citet{Liskowski2016} normalized the patches by subtracting the mean and dividing by the standard deviation of its elements.

Similar to principal component analysis (PCA) processing, the whitening processing can remove the universal correlations among the neighbor pixels of the image. 
These universal correlations are redundant for training the network. \citet{Liskowski2016} used the zero-phase component analysis to process the image data using rotations, resulting in whitened data that are as close as possible to the original data. 
Whitening pre-processing has also been used in \citep{Marques2016}.

Training a CNN for a computer-vision task requires tens of thousands of natural images. 
However, for vessel tracking, the number of training images is relatively small, which may cause an over-fitting problem.
To solve this problem, data augmentation can increase the number of training samples using image-transformation approaches (e.g., rotation, scaling, flipping, and mirroring). 
These transformations yield the desired invariance and robustness properties of the resulting network. 
The augmentation methods vary according to different tasks. \citet{Charbonnier2017} preserved the orientation of patches with four angles and obtained examples using horizontal flipping. 
In addition to the rotation and flipping used in \citep{Zhao2018f}, scaling and mirror operations were used in \citep{Huang2018} and \citep{Guo2019}, respectively. 
Moreover, random elastic deformation was used to deform the training set \citep{Fan2018, Livne2019}. 
\citet{Zreik2018a} observed the signs of over-fitting when training was implemented without data augmentation. 
The results in \citep{Lin2019} show that data augmentation is essential to achieve excellent performance.

The application of generative adversarial networks (GANs) is highly promising augmentation approach \citep{Costa2018, Yu2019}. 
By simply sampling a multi-dimensional normal distribution, \citet{Costa2018} employed the encoder-decoder network to generate realistic vessel networks and extend the training samples. \citet{Yu2019} used the shape-consistent GAN to generate synthetic images that maintain the background of coronary angiography and preserve the vascular structures of retinal vessels. 
This model can transfer the knowledge of the vessel segmentation from a public dataset to an unlabeled dataset.

\subsubsection{Sampling strategies}\label{Strategies of sampling}
There are two categories of CNNs according to the input: patch-based and image-based networks. 
The former extracts numerous patches from the image data as training samples of the network, whereas the latter considers the entire image as a training sample.

For patch-based networks, an efficient extraction strategy should be adopted. 
\citet{Nardelli2018a} extracted the patches from the CT image around the vessel of interest. 
\citet{Wolterink2019} directly selected the positive and negative samples focused on the vessel centerlines. 
Instead of extracting the patches, images can be directly inputted into the network to optimize the model; examples can be found in \citep{Hu2018, Mo2017}.
More flexible, \citet{Girard2019} used a scalable encoding-decoding CNN model that can input either the entire image directly or patches of any size to the network.

In addition to the samples directly extracted from the image, several works selected the training samples from enhanced images to focus on problems originating from vessel tracking. 
For example, \citet{Nardelli2017, Nardelli2018} extracted patches from the bronchus image enhanced by the scale-space particle approach. \citet{Hajabdollahi2018} trained the CNN on the enhanced gray-scale level image. 
\citet{Zhao2018g} selected the patches from both the original and tube-level label images. 
To directly reflect the stenosis of vessels, \citet{Zreik2018a} collected the patches from multi-planar reformatted images.

\subsubsection{Formulation of loss functions}\label{Formulation of loss functions}
The CNNs obtain the optimal network weights by optimizing the loss function. 
The cross-entropy-based loss functions are generally used in vessel-tracking tasks \citep{Dasgupta2017, Nardelli2018a, Wu2018b, Jin2019, Dharmawan2019, Mo2017,Guo2019, Lin2019}. 
However, the vascular regions in the images are considerably smaller than the non-vascular regions, thereby inducing the imbalance problem for loss-function optimization.

Data-imbalance problems occur in image segmentation, where the number of foreground pixels is usually less than background.
To resolve the imbalance problem, some researchers formulated weighted schemes for the categorical cross-entropy loss functions \citep{Li2018b, Zhang2018g, Hu2018}.
These loss functions incorporate the coefficients to reduce the importance of well-classified examples and focus on problematic samples.
Another solution that has been employed is the use of the dice-coefficient-based loss functions \citep{Livne2019, Soomro2019b, Kitrungrotsakul2019}.
To balance the classes of voxels, weighted schemes of the dice coefficient are employed to formulate the loss functions.
\citet{Huang2018} adjusted the penalty weights of misclassified voxels to obtain higher correct classification scores and lower number of misclassified voxels. 
\citet{Lian2018a} employed a tuning parameter to determine whether precision (i.e., positive prediction value) contributes more than recall (i.e., true positive rate or sensitivity) or conversely during the training procedure.

Alternative methods employ the squared error \citep{Li2016b, Fan2017} or more complex formulations as loss functions to train CNNs. 
Based on the L1 norm, \citet{Yan2018e} generated a joint loss to simultaneously train the model with the vessel-segment-level loss and pixel-wise loss.
\citet{Jiang2019} formulated the loss function by adjusting the weights of two parts: cross entropy and L2 norm.
\vspace{-1em}
\section{Evaluation issues}\label{Evaluation issues}

\subsection{Metrics for performance evaluation}\label{Assessment parameters}

%[vessel tracking的表示]
The results of vessel tracking can be presented as key points, vessel centerlines, and label images depending on requirements of the clinical applications.
The performances of the methods are evaluated by comparing the results with the ground truth.
%[评价vessel tracking 有两类指标]
The key points can be checked visually.
For the remaining two types of results, two groups of metrics are generally used: overlap metrics and classification metrics.

%[一类是重叠类的指标]
Overlap metrics are used to evaluate the similarity between the extracted vessels and ground truth.
For label images, true positive (TP), true negative (TN), false negative (FN), and false positive (FP), compared with the ground truth, are typically used to evaluate the vessel and non-vessel patterns;
these four metrics can be further formulated as accuracy (Acc), sensitivity (Se), specificity (Sp), precision (Pr), recall (Re), positive predictive value (PPV), negative predictive value (NPV), and dice similarity coefficient (DSC \citep{1945Measures}).
The DSC and the Hausdorff distance (HD) are widely adopted overlap metrics to assess the similarity between label images.
For centerlines, the four metrics can be computed according to the point-to-point correspondence between the ground truth and computed centerline. 
The four metrics can be further formulated as overlap (OV), overlap until the first error (OF), and overlap with the clinically relevant part of the vessel (OT).
The average inside (AI) distance can also be used to describe the average distance of connections between two centerlines. 
The details of the four metrics for assessing vessel centerlines are found in \citep{Schaap2009}.

%[一类是评价分类器的指标]
Classification metrics are derived from the curves (i.e., receiver operating characteristic (ROC) curve and precision-recall curve) to assess a binary classifier system.
The area under the receiver operating characteristic curve (AUC) metric can be used to indicate the probability that a classifier will rank a randomly chosen vessel instance higher than it will rank a randomly chosen non-vessel instance. 
The AUPRC metric, i.e., the area under the precision-recall curve, can also be exploited to evaluate the results of vessel tracking.
\subsection{Public datasets  and validation strategies}\label{Public datasets  and validation strategy}
%[和vessel tracking相关的一些挑战赛和数据集]

Standard datasets are required for an objective evaluation of vessel-tracking methods.
Here, we summarize the challenges and the public datasets related to vessel tracking.
\begin{enumerate}
\small
 \item[(1)] Retinal vessel segmentation: DRIVE\\ (\url{http://www.isi.uu.nl/Research/Databases/DRIVE/}) 
 \item[(2)] Retinal vessel segmentation: STARE \\(\url{http://cecas.clemson.edu/~ahoover/stare/})
 \item[(3)] Retinal vessel segmentation: CHASE-DB1 \\(\url{http:// blogs.kingston.ac.uk/retinal/chasedb1})
  \item[(4)] Retinal vessel segmentation: HRF \citep{Odstrcilik2013}
    \item[(5)] Retinal vessel segmentation: PRIME-FP20 \citep{Ding2020a}
  \item[(6)] Coronary artery stenosis detection: CASDQEF \\(\url{http://coronary.bigr.nl/stenoses})
    \item[(7)] Coronary centerline extraction: CAT08\\ (\url{http://coronary.bigr.nl/centerlines})
    \item[(8)] Identify coronary calcifications: orCaScore \\(\url{https://orcascore.grand-challenge.org/})
    \item[(9)] Coronary segmentation: ASOCA\\ (\url{https://asoca.grand-challenge.org/})
    \item[(10)] Lung vessel segmentation: VESSEL12 \citep{Rudyanto2014}
   \item[(11)] Liver segmentation: SLIVER07\\ (\url{https://sliver07.grand-challenge.org/})
    \item[(12)] Liver vessel segmentation: 3D-IRCADb \\(\url{https://www.ircad.fr/research/3d-ircadb-01/})
 \end{enumerate}

%[基于数据集选择不同的数据集划分方式（这个部分是否多余？）]
To validate the vessel-tracking methods, the dataset is divided into different training and test groups according, e.g., one-off train + test, leave-one-out and k-fold cross-validation.
The experiment columns in \Zxhreftbs{tab:traditional learning:retinal vessel} - \ref{tab:deep learning:other vessels} present the validation strategies of the selected methods.
Moreover, \Zxhreftb{tab:typical results} presents the public datasets and selected state-of-the-art results.

\begin{table*}
\caption{The public datasets and selected results. Please refer to Section \ref{Evaluation issues} for the abbreviations.}
\label{tab:typical results}\vspace{-0.2cm}
\centering
\scriptsize
\begin{tabular}{p{1.2cm}p{4cm}p{5.5cm}}
\hline\hline
\multirow{1}{*}{Public dataset}&\multirow{1}{*}{Selected Results}&\multirow{1}{*}{Validation strategy}\\
\hline
\multirow{4}{*}{DRIVE}&Acc=0.9692 \citep{Lian2019} &20 for training, 20 for testing, one-off train + test\\
{}&AUC=0.9814 \citep{Cherukuri2020}&20 for training, 20 for testing, five-fold cross-validation\\
{}&Se=0.9382 \citep{Shin2019} &20 for training, 20 for testing, one-off train + test\\
{}&Sp=0.9861 \citep{Lian2019} &20 for training, 20 for testing, one-off train + test\\
\hline
\multirow{4}{*}{STARE}&Acc=0.9740 \citep{Lian2019} &10 for training, 10 for testing, one-off train + test\\
{}&AUC=0.9882 \citep{Zhang2018g}&15 for training, 5 for testing, four-fold cross-validation\\
{}&Se=0.9598 \citep{Shin2019} &10 for training, 10 for testing, one-off train + test\\
{}&Sp=0.9916 \citep{Lian2019} &10 for training, 10 for testing, one-off train + test\\
\hline
\multirow{4}{*}{CHASE-DB1}&Acc=0.9770 \citep{Zhang2018g} &21 for training, 7 for testing, four-fold cross-validation\\
{}&AUC=0.9900 \citep{Zhang2018g}&21 for training, 7 for testing, four-fold cross-validation\\
{}&Se=0.9463 \citep{Shin2019} &20 for training, 8 for testing, one-off train + test\\
{}&Sp=0.9909 \citep{Lian2019} &21 for training, 7 for testing, four-fold cross-validation\\
\hline
\multirow{4}{*}{HRF}&Acc=0.9651 \citep{Jin2019} &15 for training, 30 for testing, one-off train + test\\
{}&AUC=0.9838 \citep{Shin2019}&15 for training, 30 for testing, one-off train + test\\
{}&Se=0.9546 \citep{Shin2019} &15 for training, 30 for testing, one-off train + test\\
{}&Sp=0.9823 \citep{Zhao2018f} &22 for training, 23 for testing, one-off train + test\\
\hline
\multirow{4}{*}{CAT08}&AI=0.21mm \citep{Wolterink2019} &7 for training, 1 for testing, leave-one-out\\
{}&OF=0.815 \citep{Wolterink2019}&7 for training, 1 for testing, leave-one-out\\
{}&OT=0.971 \citep{Schaap2011} &8 for training, 24 for testing, one-off train + test\\
{}&OV=0.969  \citep{Schaap2011} &8 for training, 24 for testing, one-off train + test\\
\hline\hline
\end{tabular}
\end{table*}

\section{Conclusion and discussion}\label{Conclusion and discussion}

We have reviewed the recent literature on vessel tracking, particularly those on the methodologies that apply machine learning, including conventional machine-learning and deep-learning algorithms.
Instead of reviewing the methods for a single application (e.g., retinal vessel segmentation and coronary centerline extraction) or based on a specific imaging modality (e.g., colored image and CTA), this paper focuses on reviewing the learning-based methods of tracking vessels of various organs in different imaging modalities.
Learning-based methods offer the advantages of mapping the input data into representative and discriminative vascular features.
Particularly, conventional learning-based methods learn the vessel-dedicated information from numerous hand-crafted features.
They can employ different classifiers to distinguish the vessels from an analogous background according to the learnt features.
Moreover, these techniques can describe the vessels with learnt vessel-dedicated parameters using statistical models.
In contrast, based on various CNN architectures, deep-learning-based methods leverage hierarchical features that can encode global and local vascular structures.

Owing to the complex morphologies of objects and image characteristics, vessel tracking is an exigent task.
Thin vessels are not observed in many vessel-segmentation tasks because of their complex structures and small sizes, and to distinguish the small-sized vessels from artifacts and noise, high-quality local textures are required.
Moreover, vessels with uncertain branches and tortuosity are difficult to track because of the complex branch connections. 
More auxiliary information (e.g., key points and orientations) should be obtained to reconstruct these vessels.
Specifically, surrounding tissues and image noise may interfere with vessel tracking because of their positions and image intensities. To reduce the interference, a series of pre-processing techniques should be considered.

The recent literature on vessel tracking mainly reports the advanced machine-learning methodologies in view of their considerable modeling capacities and potential in extracting effective features. Nevertheless, the following two problems should be considered when a new algorithm for this task is developed.

First, learning-based methods may deliver limited performance in tracking a complete vessel because of the lack of high-level vascular features (e.g., branches and connections).
The models can be trained based on hand-crafted features (Section \ref{Vessel tracking using conventional machine learning}) or hierarchical features (Section \ref{Vessel tracking based on deep learning}). 
However, in clinical practice, developing vessel-tracking methods may be exigent because of the problems involved in detecting abnormal vessels or vessels with pathologies.
To describe these vessels  in detail, the extraction of high-level features for future learning-based tracking methods is required.

The second problem is related to the strategies employed to deal with the limited training data because this insufficiency generally leads to poor generalization capacity of models.
Deep learning has achieved considerable success in many applications where  public datasets with annotation are available.
However, in the field of medical image analysis, overcoming the limitation of training data is still a major challenge.
Currently, data augmentation is a common strategy employed to alleviate this issue.
Moreover, weakly supervised and self-supervised learning are potential approaches to resolve the problems of lacking annotated data.
Hence, in the future, more public databases are expected to be available, such as via open challenges, to promote the learning-based vessel-tracking algorithms.

\section*{Acknowledgment}
This work was funded by the National Natural Science Foundation of China (grant no. 61971142 and 62011540404) and the development fund for Shanghai talents (no. 2020015)

\bibliographystyle{cas-model2-names}\biboptions{authoryear}
\bibliography{AllBibliography20190529}
\end{document}